\documentstyle[12pt,aaspp4,epsfig]{article}

\newcommand{\onee}	{\mbox{\rm\,1E~1740.7--2942}}
\newcommand{\xj}        {\mbox{\rm\,XTE~J1739--302}}
\newcommand{\dg}        {\mbox{$^{\circ}$}}
\begin{document}
\input epsf

\title{XTE J1739-302: An Unusual New X-ray Transient}

\author{D. M. Smith\altaffilmark{1}, D. Main\altaffilmark{1},
F. Marshall\altaffilmark{2}, J. Swank\altaffilmark{2},
W. A. Heindl\altaffilmark{3}, M. Leventhal\altaffilmark{4},
J. J. M. in 't Zand\altaffilmark{5}, J. Heise\altaffilmark{5}}

\altaffiltext{1}{Space Sciences Laboratory, University of California Berkeley, 
Berkeley, CA 94720}
\altaffiltext{2}{NASA Goddard Space Flight Center, Code 666, Greenbelt, 
MD 20771}
\altaffiltext{3}{Center for Astrophysics and Space Sciences, Code 0424, University
of California San Diego, La Jolla, CA 92093}
\altaffiltext{4}{Dept. of Astronomy, University of Maryland College Park,
College Park, MD 20742}
\altaffiltext{5}{Space Research Organization of the Netherlands,
Sorbonnelaan 2, 3584 CA Utrecht, The Netherlands}

\begin{abstract}

A new x-ray transient, designated \xj , was discovered with the
Proportional Counter Array (PCA) on the \it Rossi X-ray Timing
Explorer (RXTE) \rm in data from 12 August 1997.  Although it was the
brightest source in the Galactic Center region while active (about
$3.0 \times 10^{-9}$ergs cm$^{-2}$s$^{-1}$ from 2 to 25 keV), it was
only observed on that one day; it was not detectable nine days earlier
or two days later.  There is no known counterpart at other
wavelengths, and its proximity to the Galactic Center will make such
an identification difficult due to source confusion and extinction.
The x-ray spectrum and intensity suggest a giant outburst of a
Be/neutron star binary, although no pulsations were observed and the
outburst was shorter than is usual from these systems.

\end{abstract}


\section{Introduction}

\begin{figure}[t!]
\centerline{\epsfysize=2.5in \epsfbox{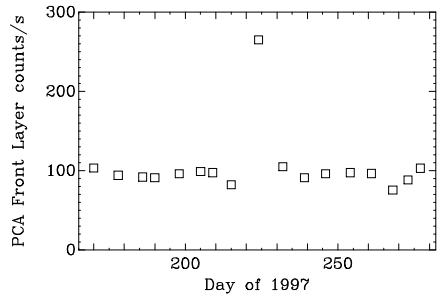}}
\caption{
Count rates from weekly PCA pointings
toward \onee .  Both instrumental and Galactic diffuse background have been
subtracted.  The high point is the 12 August observation.  Statistical
errors are smaller than the symbols used for the data points.
}
\end{figure}

Most bright x-ray transients lasting days or weeks come from
two kinds of systems: black hole binaries with a low-mass companion
(x-ray novae) and neutron star binaries with a high-mass, Be-type
companion (Be/NS).  In the latter systems, the neutron star is in a
relatively wide, often eccentric orbit, and the outburst occurs at the
orbital phase when the neutron star passes through the dense equatorial wind
of the Be star, accreting wind material.  These systems can be
distinguished from each other and from rarer kinds of transients by
neutron star pulsations (if present), by optical observations of the
companion (if available), by regularity of recurrence (which indicates
a Be/NS binary), by a weak secondary outburst peak (which is a
signature of an x-ray nova), or by their x-ray spectra (see 
sections 5 and 6).

We report a new x-ray transient in the Galactic Center region.  This
source, \xj , was discovered in an observation of the black hole
candidate \onee\ with the Proportional Counter Array (PCA) on the \it
Rossi X-ray Timing Explorer (RXTE)\rm .  The transient was initially
reported in an IAU Circular (\markcite{Sm97a}Smith et al. 1997a), with
a corrected position reported later (\markcite{Sm97}Smith 1997),
unfortunately after radio observations of the original field
(\markcite{Hj97}Hjellming 1997).

\begin{figure}[t!]
\centerline{\epsfysize=3.0in \epsfbox{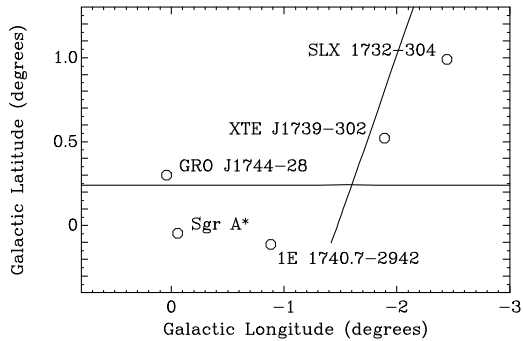}}
\caption{
Map of the Galactic center region showing well-known x-ray sources,
the two scans from 12 Aug 1998, and the resulting position fit for
XTE J1739-302. GRO J1744-28 (the bursting pulsar) and
SLX 1732-304 (an x-ray burster in Terzan 1), were not active at
this time.}
\end{figure}

On 12 August 1997 at 16:58 UT, the PCA scanned its 1\dg\ FWHM field of
toward a weekly observing position near \onee , approaching almost
exactly along a line of constant right ascension.  The count rate,
instead of rising smoothly from the background level as \onee\ entered
the field of view, instead rose much higher just before the end of the
scan, came back down, and leveled off at a rate over twice that
expected for \onee\ after background subtraction (Figure 1).  This
indicated a bright source positioned just before \onee\ in the scan
direction, but still within the field of view of the PCA once it came
to rest.

\begin{figure}[t!]
\centerline{\epsfysize=5.5in \epsfbox{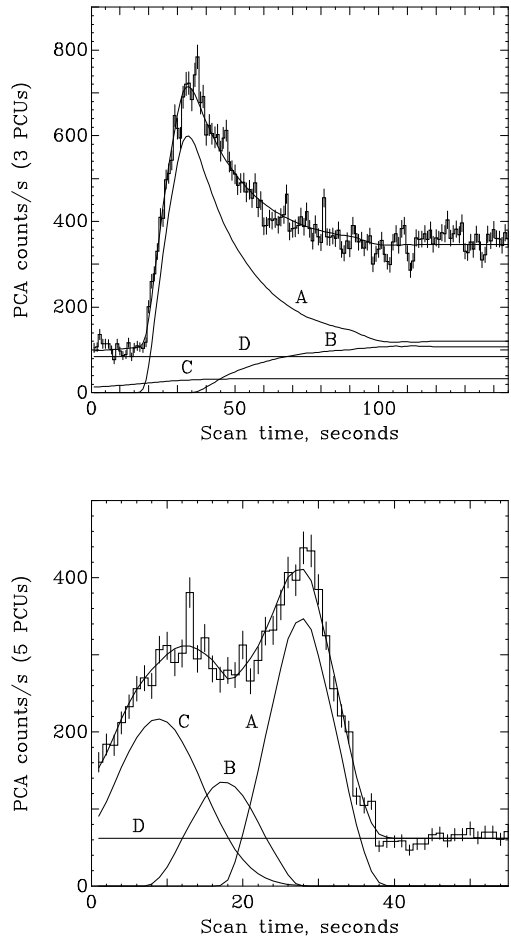}}
\caption{
PCA count rates during the two scans over \xj .  Top: the scan at
nearly constant right ascension.  Bottom: the scan at constant
Galactic latitude. The best-fit curve for each is made up of a sum of
four contributions, each in the form of a collimator response times an
intensity.  The four components are also shown and labeled as follows:
A) \xj\ in its best-fit position; B) \onee\ ; C) Galactic Center
diffuse emission (a different component in each plot - see text);
D) a constant instrumental background.  
}
\end{figure}

Two days later, on 14 August 1997, another scan passed within 0.55\dg\
of \xj , but no peak was seen.  This sets a 3$\sigma$ flux upper limit
of about 4\% of the flux on 12 August at 16:58 UT.  The limit for 3
August can be derived from Figure 1: the rms variation from week to
week in the observed flux from \onee\ (not counting the 12 August data
point) is 8 counts/s.  Since the 3 August observation is, in fact,
lower than average, the upper limit for \xj\ on this day is even
stricter than the 3$\sigma$ variation of 24 c/s, or 14\% of the 12
August flux of \xj .

Data from the Wide Field Camera on \it Beppo SAX \rm show no detection
of the source on 6 September, with an upper limit of 20 mCrab from 2
to 10 keV, roughly 30\% of the outburst flux.

\section{Source Location}

A single scan allows good determination of source position along the
scan direction, but not in the perpendicular direction.  Fortunately,
another \it RXTE \rm investigation (by F. Marshall and collaborators)
had scanned along the Galactic Plane at a Galactic latitude of
0.25\dg\ earlier the same day (14:25 UT).  Figure 2 shows the two
scans, the positions of some well-known x-ray sources,
and the position we eventually fit for the new source.  Figure 3 shows
the count rates during each scan. Together, the two scans provide a
good localization.

To find the position, we set up a grid of candidate positions spaced
0.01\dg\ apart. For each candidate position, we used the standard PCA
data analysis tool ``pcaclrsp'' to find the collimator response versus
time for each scan.  The same tool was used to find the response to
\onee .  Since the Galactic Plane emission is more complicated, we
used a different approximation for each scan.  For the scan along
right ascension, we modeled the Galactic Plane emission as decaying
exponentially in Galactic latitude with an adjustable scale height
(\markcite{Ya96}Yamasaki et al. 1996).  For the scan along Galactic
longitude, we are only concerned with the excess at the Galactic
Center compared to the rest of the plane emission
(\markcite{Ya90}Yamauchi et al. 1990). We modeled this as a Gaussian
with an adjustable FWHM.

For each scan, we fit the data with the sum of a constant background
(a good approximation) and multiples of the model lightcurves for
\onee , the Galactic emission, and the new source.  This gave four
normalization parameters for each scan.  By fitting these separately,
we allowed the sources to vary in intensity between the two scans
without affecting the final result.  Although the Galactic diffuse
emission is not variable, it also had to be fit separately because it
was treated differently for the two scan directions.  Repeating this
process for each point on the grid of candidate positions, we found
the best-fit position for the new source to be right ascension
$17^{\rm h} 39^{\rm m} 00^{\rm s}$, declination $-30$\dg\ 16'.2
(J2000).

The 99\% statistical confidence interval for this fit is an oval with
radii 1.5 arcmin in right ascension and 2.2 arcmin in declination
(both in true minutes of arc, and calculated assuming two parameters
of interest).  There are, however, two sources of systematic error:
the modeling of the diffuse emission, and intensity fluctuations in
the source.

\begin{figure}[t!]
\centerline{\epsfysize=2.5in \epsfbox{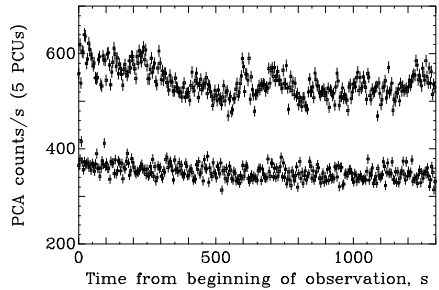}}
\caption{
Raw count rates in 4 s intervals during the
pointings toward \onee\ on 12 August (top, including \xj ) and 20
August (bottom, typical of all \onee\ pointings).  Unlike Figure 1,
these rates include instrumental and Galactic diffuse background.  
}
\end{figure}

As modeled, the FWHM of the Galactic Center Gaussian was 1\dg\ and the
scale height of the Galactic Plane exponential was 3\dg .  The
best-fit position of \xj\ is very insensitive to these parameters, and
varies by less than 1 arcmin even as the Gaussian width varies from
0\dg\ to 2\dg\ and the exponential scale height varies from 1\dg\ to
5\dg .

Because the source shows low-frequency noise (see below), a
fluctuation of the same timescale as the scan across the source could
distort its apparent position.  To estimate the size of this effect,
we used the relatively long (1300 s) stretch of data after the scan to
\onee\ had reached its observing position (see Figure 4) to
represent the likely ``population'' of fluctuations.  This lightcurve
was smoothed to eliminate the high-frequency (mostly statistical)
noise, and sampled at random positions to provide percentage
fluctuations which could be introduced to each of the two short stretches of
scan data.  This process was repeated many times to get a
distribution of best fit positions.  The contour which encloses 99\%
of these trials has radii 4.5 arcmin in right ascension and 4.2 arcmin in
declination, showing that this is the dominant source of uncertainty.
Adding in quadrature the three contributions from statistics, modeling, and
fluctuations, we find an overall 99\% confidence
contour which is almost perfectly circular, with a radius of 4.8 arcmin.

\section{Luminosity}

The luminosity of \xj\ is also obtained from the fits for each scan.
For the scan at 16:58 UT, the counting rate at the peak of the collimator
response to \xj\ is 600 c/s using three of the five detectors of the
PCA (the other two were not turned on until the scan ended).  For the
best fit position, this represents a collimator response of 91\%, thus
the PCA count rate for all five detectors pointed directly at the
source would have been 1100 c/s.  During the Galactic Plane scan only
2.55 hours earlier, the flux was only 56\% of this value.  We
cannot know if we caught the source turning on or if it
had large-amplitude variations during a longer active period.

Assuming a thermal bremsstrahlung spectrum (see below), the x-ray flux
from 2.5 to 25 keV is $3.0 \times 10^{-9}$ ergs cm$^{-2}$s$^{-1}$.
Removing the effect of the absorption column, this becomes $3.6 \times
10^{-9}$ergs cm$^{-2}$s$^{-1}$, or $2.1 \times 10^{-9}$ergs
cm$^{-2}$s$^{-1}$ from 2 to 10 keV and $4.8 \times 10^{-9}$ergs
cm$^{-2}$s$^{-1}$ from 2 to 100 keV.  Both the position and absorption
column (see below) support a location near the Galactic Center, so
assuming a distance of 8.5 kpc, the 2 to 100 keV luminosity is $4.2
\times 10^{37}$ergs s$^{-1}$.

\section{Timing}

\begin{figure}[t!]
\centerline{\epsfysize=3.0in \epsfbox{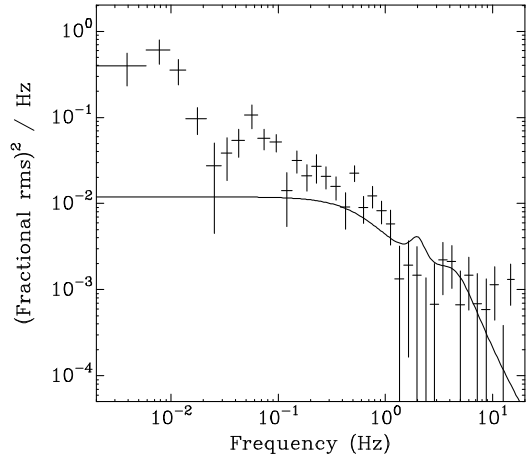}}
\caption{
Power spectrum of the 12 August observation. Fractional rms is
expressed relative to the count rate due to \xj\ alone.  Poisson
noise has been subtracted.  The curve shown is the power spectrum
of \onee\ from Smith et al. (1997b) rescaled
so that it could be directly subtracted.
}
\end{figure}

Figure 4 shows the lightcurve of the 12 August \onee\ observation once
the scan ended, with the 20 August observation shown for comparison.
It is apparent that the source contributing the additional flux has
significant low-frequency noise.  Figure 5 shows the 12 August power
spectrum along with a fit to the power spectrum from a long
observation of \onee\ alone (\markcite{Sm97b}Smith et al. 1997b).  The
total rms variability of \xj , removing the effects of Poisson
statistics and the usual variability of \onee , is 13\%, integrated
from 0.003 to 10 Hz.  We have examined the power spectrum with fine
frequency resolution from 0.01 to 1000 Hz and find no statistically
significant periodicities or QPOs.  We can exclude with at least 99\%
confidence pulsations with an amplitude above 2\% for any period
shorter than 300 s.

\section{Spectrum}

\begin{figure}[t!]
\centerline{\epsfxsize=3.3in \epsfbox{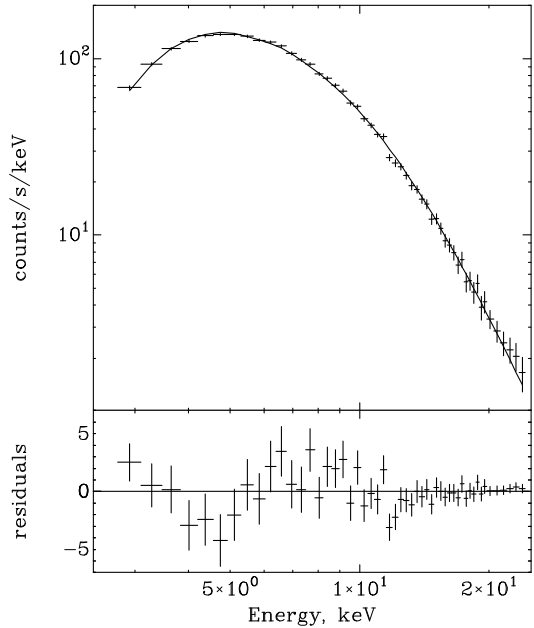}}
\caption{
\xj\ count spectrum (\onee\ and background subtracted), 
best fit thermal bremsstrahlung spectrum, and residuals from XSPEC.
The collimator response has been removed, so this is the equivalent
on-axis intensity.
}
\end{figure}

We were able to remove the spectral contributions from Galactic diffuse
emission and \onee\ with high confidence by subtracting a spectrum
constructed from six of the pointings shown in Figure 1, three before
and three after the 12 August event.  Both the intensity and spectral
shape of \onee\ have been quite stable over the last two years.

Despite the lack of pulsation, the energy spectrum of \xj\ is well fit
by the models which have been historically used for x-ray
pulsars.  The one most commonly used (e.g. \markcite{Wh83}White, Swank
\& Holt 1983) has been a power law with energy index $\alpha$ and an
exponential cutoff (i.e. $E^{-\alpha-1} {\rm exp}[(E_c - E)/E_f]$ ph
cm$^{-2}$ s $^{-1}$ above $E_c$ and just the power law below $E_c$).
We find $\alpha = (0.38 \pm 0.12)$ (i.e. a photon index of $-1.38$),
$E_c = (6.6 \pm 0.8)$ keV, and $E_f = (21.4 \pm 4.5)$ keV.  The very
low reduced $\chi^2$ (0.78) indicates that five parameters may not be
needed.  Figure 6 shows the count spectrum from 2.5 to 25 keV and an
optically thin thermal bremsstrahlung (OTTB) model from XSPEC
(\markcite{Ar96}Arnaud 1996) which has only three parameters
(temperature, normalization, and absorption column).
The XSPEC model is taken from \markcite{Ke75}Kellogg, Baldwin \& Koch
(1975).  The fit is very good ($\chi^2_r$ = 1.075 for 49 degrees of
freedom).  The fit temperature is (21.6 $\pm$ 0.8) keV.  A common
approximation to a bremsstrahlung spectrum at high energies is
$E^{-1}{\rm exp}[-E/kT]$.  This also fits well
($\chi^2_r$ = 1.054), giving $T = $(12.4 $\pm$ 0.3) keV.
Each model included an absorption column; the values were (5.3 $\pm$
0.7), (6.2 $\pm$ 0.2) and (5.1 $\pm$ 0.2) $\times 10^{22}
\rm{cm}^{-2}$, respectively, consistent with the position
near the Galactic Center.

Other spectral forms can be excluded.  A power law is a poor fit
($\chi^2_r$ = 2.347), as is a multicolored disk model ($\chi^2_r$ =
3.072).  We also tried the sum of these two, as representative of a
typical black hole spectrum.  The fit is good ($\chi^2_r$ = 0.674),
but this is to be expected when using a five-parameter model for a
spectrum which was fit well by a three-parameter model (OTTB).  The
resulting parameters are not typical of black hole candidates: the
power law photon index (1.0) and disk blackbody temperature (3.3 keV)
are both harder than any ever seen.  We therefore do not consider this
fit physically meaningful.

\section{Discussion}

Most bright Galactic x-ray transients are either black hole candidates
or binaries consisting of a Be star and a neutron star (Be/NS).  We
tentatively identify \xj\ as a Be/NS binary because its spectral shape
is similar to that of these systems: a gradual steepening over the 2
to 25 keV range.  In this energy range, the black hole candidates show
either two distinct components (a very soft thermal component and a
power law) or else a single, very hard power law.  Our parameters for
the cutoff-power-law model of the \xj\ spectrum fall within the ranges
of the x-ray pulsars listed in \markcite{Wh83}White et al. (1983),
which include Be/NS binaries.  Our temperature for the exponential
approximation to a bremsstrahlung spectrum (12.4 keV) is near the low
end of the ranges of GS~0834-430 (12-18 keV; \markcite{Wi97}Wilson et
al. 1997) and A0525+262 (14 to 22 keV; \markcite{Fi96}Finger, Wilson
\& Harmon 1996) measured with BATSE.  As noted above, spectral forms
typical of black hole candidates were either poor fits or gave
parameters far from the usual range.

With the OTTB model used in section 5, the luminosity from 20 to 50
keV $(1.05 \times 10^{37}$ergs s$^{-1}$, about 1/4 of the 2 to 100 keV
luminosity) is at the high end of the expected range for a ``giant''
outburst of a Be/NS binary (\markcite{Bi97}Bildsten et al. 1997).  The
short duration ($<$ 11 days) would be not be typical for a
giant outburst (\markcite{Bi97}Bildsten et al. 1997), although one
giant outburst of A0535+262 was only 18 days long
(\markcite{Se90}Sembay et al. 1990).  With no detected pulsations and
such a short sample of data, we can't distinguish between a normal and
giant outburst by spin-up rate.  

If we caught the source turning on, then at $< 2$ days it is not
similar to any common class of transient, although the lightcurve is
consistent with that recently observed from XTE J0421+560 = CI Cam
(\markcite{Sm98}Smith et al. 1998).  That bright transient showed a
fast rise over hours followed by an exponential decay with a time
constant on the order of a day (quick-look results provided by the
ASM/\it RXTE \rm team).  Otherwise, it behaved very differently from
\xj , showing a bright Fe-K line and no time variability besides its
smooth decay (\markcite{Be98}Belloni et al. 1998).  What the events
have in common is that they underscore the need for very frequent
monitoring and very fast response in the study of Galactic x-ray
transients.

If we are correct in the spectral identification of \xj\ as a Be/NS
binary, the pulse period may simply have been too long ($>$ 300 s) to
be seen in our short (1300 s) stretch of data, but the lack of
detection could also be due to a very low pulsed fraction.  A pulsed
fraction of only 4 to 6\% rms was seen in X0331+53 (= V0332+53 = BQ
Cam) along with flat-topped noise with a total rms of 25\% from 0.01
to 1 Hz (\markcite{Ma90}Makishima et al. 1990).  Our power spectrum is
similar to that of the A0535+262 outburst of 1980 October with its
pulsed component removed (\markcite{Fr85}Frontera et al. 1985).
Although that power spectrum had a broad peak around 0.05 Hz detected
with more significance than the similar feature in our spectrum (see
Figure 5), the authors interpreted it as due to imperfect subtraction
of the pulsation, so the coincidence should be viewed cautiously.
Like ours, their power spectrum showed significant low-frequency noise
below 0.02 Hz and little power above 0.1 Hz.

Our best hope to learn more about this object is to catch a subsequent
outburst with a longer exposure and multiple instruments.  If it is a
Be/NS binary, its unusually brief and luminous outburst makes
particularly interesting.  We encourage anyone with access to
appropriate archival data to look for the August 1997 outburst and any
earlier ones, particularly to search for pulsar periods longer than
300 s.  We also encourage anyone with new Galactic Center data to look
at it promptly and call for target of opportunity observations from
the various high-energy observatories during any subsequent outburst
of this object.

The authors would like to thank the \it RXTE \rm Guest Observer
Facility for their continuing assistance, and R. Hjellming for his
prompt response in scheduling radio observations of the originally
reported position.  This work was supported by NASA grants NAG5-3599
and NAG5-4110.

\end{document}